# Introduction of deep level impurities, S, Se, and Zn, into Si wafers for high-temperature operation of a Si qubit

Yoshisuke Ban[1]*, Kimihiko Kato[2], Shota Iizuka[2], Shigenori Murakami[2], Koji Ishibashi[1,3], Satoshi Moriyama[4], Takahiro Mori[2], and Keiji Ono[1,3]*

[1]*Advanced device laboratory, Cluster for Pioneering Research (CPR), RIKEN, Wako, Saitama 351-0198, Japan*
[2]*Device Technology Research Institute (D-Tech), National Institute of Advanced Industrial Science and Technology (AIST), Tsukuba, Ibaraki 305-8568, Japan*
[3]*Center for Emergent Matter Science (CEMS), RIKEN, Wako, Saitama 351-0198, Japan*
[4]*Tokyo Denki University (TDU), Adachi, Tokyo 120-8551, Japan*

E-mail: yoshisuke.ban@riken.jp, k-ono@riken.jp

To realize high-temperature operation of Si qubits, deep impurity levels with large confinement energy, which are hardly thermally excited, have been introduced into Si wafers. Group II impurity Zn and group VI impurities S and Se, which are known to form deep levels, were introduced into the Si substrates by ion implantation. These samples were analyzed for concentration-depth profiles, energy level depths, and absence of defects. To introduce deep impurities into thin channels such as 50-nm-thick Si, we found impurity introduction conditions so that the concentration depth profiles have maximum value at less than 50 nm from the Si surface. Then, the formation of the deep levels and absence of defects were experimentally examined. By using the conditions to introduce deep impurities into Si wafer obtained from the experiments, single-electron transport at room temperature, high-temperature operation of qubit, and room-temperature quantum magnetic sensors are promising.

# 1. Introduction

Many studies on silicon qubit have been conducted worldwide due to their potential to be highly consistent with integrated circuit technology. Operation of silicon qubits has been reported at extremely low temperatures such as 0.1K.[1] If the operation temperature of qubit can be increased, it is expected to accelerate research and extend applications of the technology. The depth of deep impurity levels of S, Se, and Zn in silicon are 10 – 20 times larger than the room temperature energy (26 meV), and electrons trapped in the deep levels are hardly thermally excited. It would be possible to realize high-temperature operation of silicon qubit by introducing deep levels on purpose into silicon devices. Then, we are studying to realize qubit and quantum sensors utilizing deep impurity levels in silicon.

In a previous study, single-electron transport up to 300 K and electron spin resonance and qubit operation (Rabi oscillation) at 10 K in Al-N isoelectronic trap-introduced Si tunnel FETs[2] have been reported.[3] By introducing deep impurity levels into the channel of a tunnel FET, *i.e.* a gated p-i-n structure, single electron transport through deep levels can be realized. Furthermore, spin blockades[4] via double quantum dot formed by two deep levels enables the readout of single electron spin states and the observation of electron spin resonance and Rabi oscillation of the spin qubit. In this previous study,[5] the deep impurity level of Al-N and the shallow acceptor level were presumed to compose the double quantum dot. The reason that the operation temperature of the qubit is limited to 10 K can be interpreted as the shallow levels are subject to thermal excitation. To increase the operation temperature of the qubit, it is necessary to realize double quantum dot structure that consists of two deep impurity levels. Thus, we focused on group II-VI elements that form deep impurity levels in silicon. In our previous study,[6] we introduced Be into Si wafers and confirmed the formation of impurity levels by photoluminescence, and conducted experiments to introduce Be into Si devices.

The group II element Zn and group VI elements S and Se have been introduced into Si by diffusion or ion implantation (I/I), and the formation of deep donors and acceptors have been reported by many groups.[7]-[17] Note that Zn is a double acceptor, and S and Se are double donors, each of which forms two levels in the Si band gap. However, I/I and post-implantation annealing (PIA) conditions that realize concentration-depth profiles having maximum values at less than 50 nm from the Si wafer surface, which are compatible with Si CMOS technology, have not been reported. We have conducted this study, as in our previous work, for the purpose of introduction of deep impurity levels into a 50-nm-thick Si-on-insulator (SOI) layer and process integration into tunnel FET devices. In this study, we introduced S, Se, and Zn into Si substrates by I/I, investigated PIA conditions, and examined the formation of deep impurity levels and the presence of defects. To distinguish deep impurity levels from defect levels in the study, it is preferable to suppress defect formation with the introduction of the impurities. Therefore, the formation of deep impurity levels and the presence of defects were examined by multiple analysis methods. We found fabrication conditions to introduce deep impurity levels into Si wafers without defect formation and realized a concentration depth profile with the highest impurity concentration at less than 50 nm from the wafer surface.

This paper relates to the presentation held at the International Conference on Solid State Devices and Materials (SSDM) 2022[18]) and discusses the optimization of I/I and PIA conditions to form deep impurity levels in Si wafers by introducing S, Zn, and Se, confirmation of the formation of deep levels, and analysis of the presence of defects, in detail.

## 2. Experimental methods

Ion implantations (I/I) of $^{32}$S, $^{64}$Zn, and $^{80}$Se into Si wafers were carried out and their concentration-depth profiles and the presence of defects were examined. Here, we used CZ P-doped (n-type) Si(100) wafers with a resistivity of 5 – 10 Ω·cm or B-doped (p-type) Si(100) with a resistivity of 10 – 20 Ω·cm. In the following, $^{32}$S, $^{64}$Zn, and $^{80}$Se are denoted as S, Zn, and Se, respectively. The I/I energy and dose for S, Zn, and Se are 15 keV and $1\times10^{13}$ cm$^{-2}$, respectively. PIA was performed by flash lamp annealing (FLA) at a peak temperature of 1200°C for 1.5 to 20 msec. For these samples, concentration-depth profiles were measured by secondary ion mass spectrometry (SIMS). Then, the presence of defects was examined by cross-sectional TEM observation and photoluminescence (PL) measurements.

Subsequently, CC-DLTS (Constant Capacitance Deep Level Transient Spectroscopy) analysis was performed to observe the deep impurity levels of S, Zn, and Se. The fabrication process of planar Schottky diodes (500 μm diameter) for CC-DLTS analysis is as follows: First, S, Se, and Zn were introduced by I/I into Si wafers, respectively. The I/I energy and dose are 15 keV and $1\times10^{13}$ cm$^{-2}$, respectively. For S and Se I/I, n-type Si(100) wafers (5 – 10 Ωcm) were used, and for Zn implantation, p-type Si(100) wafer (10 – 20 Ωcm) was used. Next, PIA by FLA was performed at 1200°C for 3 msec for the S-implanted wafer and 1200°C for 1.5 msec for the Se and Zn-implanted wafers in N$_2$ gas. Next, metal electrodes were formed by sputtering and reactive ion etching to form Schottky barriers. Here, Al was used for the n-type Si wafers and Ti was used for the p-type Si wafer for the metal electrodes. Next, Al 300 nm was sputtered on the backsides of the wafers as electrodes. Finally, annealing was performed at 450°C for 30 min in 3% H$_2$/N$_2$ gas. CC-DLTS analysis of the Schottky diodes was performed with Phystech FT1030.

## 3. Results and discussion

3.1 Results

Fig. 1(a) presents the concentration-depth profiles of S obtained by SIMS analysis for Si(100) wafers with S I/I and PIA at a peak temperature of 1200°C for 3 to 20 msec, and with S and Zn I/I and PIA at 1200°C for 1.5 msec by FLA. S areal density obtained from the depth concentration profile of the as-implanted one (black line in Fig. 1(a)) is $1.1\times10^{13}$ cm$^{-2}$, which is consistent with the implantation dose $1\times10^{13}$ cm$^{-2}$. As shown in Fig. 1(a), the S areal density decreases with increasing FLA time compared to the as-implanted one: the S areal density is $1.6\times10^{12}$ cm$^{-2}$ at 3 msec (orange line in Fig. 1(a)) and $4.4\times10^{11}$ cm$^{-2}$ at 20 msec (purple line in Fig. 1(a)). This result suggests S desorption by FLA, and the longer the time of FLA, the larger the amount of S desorption. To form double quantum dots with deep levels in a p-i-n structure, we also prepared a Si wafer with both S and Zn I/I. The wafer with both S and Zn I/I and 1.5 msec FLA has an S areal density of $1.4\times10^{12}$ cm$^{-2}$ (red line in Fig. 1(a)), comparable to that of 3 msec FLA. On the other hand, the value of the S concentration peak was $\sim 3\times10^{18}$ cm$^{-3}$ for the 1.5 msec wafer with S and Zn I/I, which is higher than that of $\sim 1\times10^{18}$ cm$^{-3}$ at 3 msec. The condition of FLA at 1200°C and 1.5 msec, which results in the closest profile to the Al-N concentration profile obtained in the Al-N implanted TFETs in the previous study,[3] is mainly employed in the following experiments. Fig. 2(b) shows the concentration-depth profiles of S and Zn in the Si(100) wafer after S and Zn I/I (15keV, $1\times10^{13}$ cm$^{-2}$) and the FLA, and that of $^{80}$Se in the Si(100) wafer after $^{80}$Se I/I (15keV, $1\times10^{13}$ cm$^{-2}$) and the FLA. Here, the FLA was performed at 1200°C and 1.5 msec for the both wafers. The concentration-depth profiles of S, Zn, and Se in Fig. 2(b) are similar with a maximum concentration of $\sim 2$ to $3\times10^{18}$ cm$^{-3}$ at depths of less than 10nm. From these results, appropriate I/I and PIA condition for the introduction of S, Zn, and Se impurities into Si channels less than 50 nm thick was obtained.

Fig. 2 shows cross-sectional TEM images of (a) S and Zn implanted and (b) $^{80}$Se implanted Si(100) wafers. S, Zn, and Se I/I were conducted at 15 keV and $1\times10^{13}$ cm$^{-2}$, and FLA was conducted at 1200°C for 1.5 msec in the both wafers. The insets in Fig. 2(a) and (b) show magnified views of the respective TEM images. No apparent amorphization or defects such as {311} defects were observed in the S and Zn-implanted and Se-implanted wafers as shown in Fig. 2(a) and (b).

Fig.3 (a) – (c) present CC-DLTS spectra measured for the Schottky diodes with (a) S, (b) Se, and (c) Zn I/I, respectively. Here, n-type Si(100) wafers (5 – 10Ωcm) are used for S and Se implanted diodes, and p-type Si(100) wafer (10 – 20Ωcm) is used for Zn implanted diodes.

S, Se, and Zn I/I were performed at 15 keV and $1\times10^{13}$ cm$^{-2}$ into the Si wafers. Then, FLA at 1200°C was conducted for 3 msec for S-implanted wafers and 1.5 msec for Se- and Zn-implanted wafers. Carrier densities at 300 K obtained by C-V measurements were $9\times10^{14}$ cm$^{-3}$, $1.1\times10^{15}$ cm$^{-3}$, and $1.7\times10^{15}$ cm$^{-3}$ for S, Se, and Zn implanted diodes, respectively. The spectra of the CC-DLTS signal (mV) at period widths $T_W$ = 19.2, 192, and 1920 ms are shown in Fig. 3 (a) – (c). DLTS[19] is a method to measure the capacitance transient at constant voltage and is analyzed using an approximation under the condition that the trap density is less than the carrier density. On the other hand, CC-DLTS[20)-22)] is a method to measure voltage transients with constant capacitance and can be analyzed even if the trap concentration is larger than the carrier density. The concentration-depth profiles of S, Se, and Zn obtained by SIMS (Fig. 1) have maximum concentrations of about $1\times10^{18}$ cm$^{-3}$, which is larger than the carrier density value of $\sim1\times10^{15}$ cm$^{-3}$. Therefore, the CC-DLTS method was used in this study. CC-DLTS spectra in Fig. 3 (a), (b), and (c) were measured with a constant bias capacitance of 15pF, a voltage pulse width of 0.1 msec, and pulse heights of (a) 3.5V, (b) and (c) 2.0V. CC-DLTS peaks E1 and E2 were observed for the S- and Se-implanted Schottky diodes in Fig. 3(a) and (b), and CC-DLTS peaks H1 and H2 were observed for the Zn-implanted diodes in Fig. 3(c). A tail seen on the low-temperature side in Fig. 3(a) is not discussed here, since it is presumed to originate from a shallow level.

Fig.3(d) – (f) present Arrhenius plots for the (d)S, (e)Se, and (f)Zn implanted Schottky diodes. In these figures, $\tau$ and $T$ represent the time constant of the thermal emission of electrons and temperature, respectively. These plots were obtained from CC-DLTS spectra obtained by the correlation function method[23)-26)]. Deep double donor levels were observed at 0.25 eV (E1) and 0.50 eV (E2) for the S-implanted diode in Fig. 3(d), and at 0.23 eV (E1) and 0.46 eV (E2) for the Se-implanted diode in Fig. 3(e). For the Zn-implanted diode, deep double acceptor levels were observed at 0.26 eV (H1) and 0.62 eV (H2) in Fig. 3(f). From these results, the formation of deep impurity levels was confirmed for the Si wafers with the S, Se, and Zn I/I and FLA. In particular, no unexpected defect levels were observed for the Zn and Se-implanted Si wafers. The activation energies $E_C - E$ or $E - E_V$, and carrier capture cross sections $\sigma_n$ or $\sigma_p$ obtained from the Arrhenius plots in Fig. 3(d) – (f) are shown in Table. I. Trap concentration $N_T$ (cm$^{-3}$) for each level which is averaged over the temperature range of the Arrhenius plots is also shown in Table I. Trap concentration at each temperature is obtained from the CC-DLTS measurements as $2\Delta VC^2/e\varepsilon$, where $\Delta V$ is the amplitude of the constant capacitance voltage, $C$ is the constant capacitance per unit area, $e$ is the elementary charge, and $\varepsilon$ is the dielectric constant of Si.

Fig. 4 shows PL spectra from Si(100) wafers with (a) both Se and Zn I/I, and (b)Se I/I at 20 K. These wafers were subjected to S, Zn, and Se I/I at 15keV and $1\times10^{13}$ cm$^{-2}$, and PIA at 1200°C for 1.5 msec. The 1.132-eV, 1.093-eV, and 1.028-eV lines in Fig. 4 originated from the TA, TO, and TO+O$^\Gamma$ phonon of Si, respectively.[27] All peaks observed in Fig. 4 originated from Si only, and no other peaks were observed. Although emission lines due to S,[28),29] Zn,[30)-32] and Se[33] were reported in some previous studies, they were not observed in Fig. 4. Differences of S, Zn, and Se introduction and PIA conditions may affect the results. Defects-related emission lines[36),6] such as X-line (1.018eV)[34] and W-line (1.040eV),[35] which have been observed in ion-irradiated Si, were also not observed. These results from Fig. 1 – 4 confirm that the formation of deep impurity levels in the Si wafers was realized without the formation of unintended defect levels.

## 3.2 Discussion

We introduced S, Zn, and Se by I/I at 15 keV and $1\times10^{13}$ cm$^{-2}$ into Si(100) wafers and with PIA by the FLA, and confirmed the formation of deep impurity levels by CC-DLTS analysis (Fig.3), respectively. Furthermore, S, Zn, and Se were successfully introduced at a concentration of $\sim1\times10^{18}$ cm$^{-3}$ into the depth range of ~10 nm from the silicon surface (Fig. 2). The data in Fig. 1(a), Fig. 2(b), and Fig. 4(a) suggest that both S and Zn I/I does not interfere or change the concentration profiles or form new defects. From these results, we have clarified the conditions to introduce the deep impurity levels into Si devices compatible with Si CMOS technology. By the introduction of deep impurity levels into Si devices under the conditions of I/I and PIA obtained from this study, a high-temperature qubit could be realized. By introducing two deep impurity levels into a p-i-n structure in a Si device with the procedure obtained in this study, a spin state can be read out via spin blockade.[3] By using this technique, high-temperature Si qubit and room-temperature quantum sensors devices would be realized. No defects were observed in the cross-sectional TEM (Fig. 2), PL analysis (Fig. 4) for the S/Zn and Se-implanted Si(100) wafers, and the CC-DLTS analysis for the Se and Zn-implanted Si diodes (Fig. 3(b) and (c)). These results suggest that deep levels of Se and Zn can be introduced into Si wafers without defects. In the Schottky diodes used in this study with a carrier concentration of $\sim1\times10^{15}$ cm$^{-3}$, the depth of impurities from the metal/Si interface that can be measured by the CC-DLTS method is about 200nm – 1μm. To realize the high-temperature operation of Si qubits and quantum sensor devices with deep impurity levels in the future, we plan to introduce the two deep impurity levels into tunnel FETs with 50 nm thick SOI layers. However, we have not examined by DLTS analysis the existence of defects or impurity complex levels such as S-Zn and Se-Zn in the range of 50 nm from the surface when two impurities, such as S and Zn or Se and Zn, are introduced. Hence, further investigation would be needed to examine this issue.

## 4. Conclusions

To form deep impurity levels in Si(100) wafers, we have introduced S, Zn, and Se by I/I and PIA was performed by FLA. By SIMS analysis, I/I and PIA conditions were found with a concentration maximum (~2 to $3\times10^{18}$ cm$^{-3}$) at less than 10nm from the Si surface. Then, we confirmed the formation of deep donor and acceptor levels of S, Se, and Zn in Si wafers by CC-DLTS analysis. No defects were observed for Zn and Se-implanted Si(100) wafers in the DLTS analysis of the high sensitivity. In addition, no defects were observed in S/Zn and Se-implanted Si (100) wafers by TEM observation and PL analysis. From these results, we have successfully obtained the conditions to introduce deep impurity levels of S, Zn, and Se into Si wafers such as SOI wafers with thin Si layers (~50 nm). By introducing S, Zn, and Se into tunnel FETs with the conditions found in this study, high-temperature operation of Si qubit, and room-temperature quantum magnetic sensors can be realized in the future.

## Acknowledgments

This study was partly supported by JST CREST Grant Number JPMJCR1871 and MEXT Quantum Leap Flagship Program (Q-LEAP) Grant Number JPMXS 0118069228, Japan. A part of this work was conducted at the AIST Nano-Processing Facility, supported by "Nanotechnology Network Japan" of the MEXT, Japan.

## References

1) Zwanenburg, F. A. *et al*. Silicon quantum electronics. Rev. Mod. Phys. **85**, 961 (2013).
2) T. Mori, H. Asai, J. Hattori, K. Fukuda, S. Otsuka, Y. Morita, S. O'uchi, H. Fuketa, S. Migita, W. Mizubayashi, H. Ota, and T. Matsukawa,, IEDM Tech. Dig. (2016) p. 512.
3) K. Ono, T. Mori, and S. Moriyama, Sci. Rep. **9**, 469 (2019).
4) K. Ono, D.G. Austing, Y. Tokura, and S. Tarucha, Science **297**, 1313 (2002).
5) G. Pensl, G. Roos, C. Holm, and P. Wagner, Mater. Sci. Forum **10-12**, 911-916 (1986).
6) Y. Ban, K. Kato, S. Iizuka, S. Moriyama, K. Ishibashi, K. Ono, and T. Mori, Jpn. J. Appl. Phys. **60**, SBBA01 (2021).
7) H. G. Grimmeiss, E. Janzen, and B. Skarstam, Journal of Applied Physics **51**, 3740 (1980).
8) K. Gwozdz, V. Kolkovsky, J. Weber, A. A. Yakovleva, Y. A. Astrov, Phys. Status Solidi A, **216**, 1900303 (2019).
9) F. Richou, G. Pelous, and D. Lecrosnier, Appl. Phys. Lett. **31**, 525 (1977).
10) M. Kleverman and H. G. Grimmeiss, Phys. Rev. B **31**, 3659 (1985).
11) A.C. Wang, L.S. Lu, C.T. Sah, Phys. Rev. B **30**, 5896 (1984)..
12) S. Weiss, R. Beckmann, and R. Kassing, Appl. Phys. A **50**, 151 (1990).
13) H. Bracht, N.A. Stolwijk, and H. Mehrer, Phys. Rev. B, Phys. Rev. B **52**, 16542 (1995).
14) S. Voss, H. Bracht, and N. A. Stolwijk, Appl. Phys. Lett. **73**, 2331 (1998).
15) S. Voss, N. A. Stolwijk, and H. Bracht, Phys. Rev. B **68**, 035208 (2003).
16) A Masuhr, H Bracht, N A Stolwijk, H Overhof, and H Mehrer, Semicond. Sci. Technol. **14** 435 (1999).
17) N. Achtziger and W. Witthuhn, Phys. Rev. Lett. **75**, 4484 (1995).
18) Y. Ban, K. Kato, S. Iizuka, S. Murakami, S. Moriyama, K. Ishibashi, T. Mori, and K. Ono, Ext. Abstr. Solid State Devices and Materials, 2022.
19) D. V. Lang, J. Appl. Phys. **45**, 3023 (1974).
20) G. Goto, S. Yanagisawa, O. Wada, and H. Takanashi, Jpn. J. Appl. Phys. **13**, 1127 (1974).
21) N. M. Johnson, D. J. Bartelink, R. B. Gold, and J. F. Gibbons, J. Appl. Phys. **50**, 4828 (1979).
22) N. M. Johnson, J. Vac. Sci. Technol. **21**, 303 (1982).
23) G. L. Miller, L. V. Ramirez, and D. A. Robinson, J. Appl. Phys. **46**, 2638 (1975).
24) L. C. Kimerling, IEEE Trans. Nucl. Sci. **NS-23**, 1497 (1976).
25) Y. Tokuda, N. Shimizu, and A. Usami, Jpn. J. Appl. Phys. **18**, 309 (1979).
26) A. A. Istratov, J. Appl. Phys. **82**, 2965 (1997).
27) G. Davies, Physics Reports 176, **83** (1989).

28) D. J. S. Beckett, M. K. Nissen, and M. L. W. Thewalt, Phys. Rev. B **40**, 9618 (1989).

29) A. Yang, M. Steger, M. L. W. Thewalt, M. Cardon, H. Riemann, N. V. Abrosimov, M. F. Churbanov, A. V. Gusev, A. D. Bulanov, I. D. Kovalev, A. K. Kaliteevskii, O. N. Godisov, P. Becker, H. -J. Pohl, J. W. Ager III, E. E. Haller, Physica B **401**, 593 (2007).

30) M. O. Henry, J. D. Campion, K. G. McGuigan, M. L. W. Thewalt, and E. C. Lightowlers, Materials Science and Engineering: B **4**, 201 (1989).

31) M. O. Henry, J. D. Campion, K. G. McGuigan, E. C. Lightowlers, M. C. do Carmo, and M. H. Nazare, Semicond. Sci. Technol. **9** 1375 (1994).

32) S. E. Daly, E. McGlynn, M. O. Henry, J. D. Campion, K. G. McGuigan, M. C. do Carmo, and M. H. Nazare, Materials Science and Engineering: B **36**, 116 (1996).

33) P. L. Bradfield, T. G. Brown, and D. G. Hall, Phys. Rev. B **38**, 3533 (1988).

34) P. K. Giri, S. Coffa, and E. Rimini, Appl. Phys. Lett. **78**, 291 (2001).

35) P. K. Giri, Semicond. Sci. Technol. **20**, 638 (2005).

T. Mori, Y. Morita, and T. Matsukawa, AIP Advances **8**, 055024 (2018).

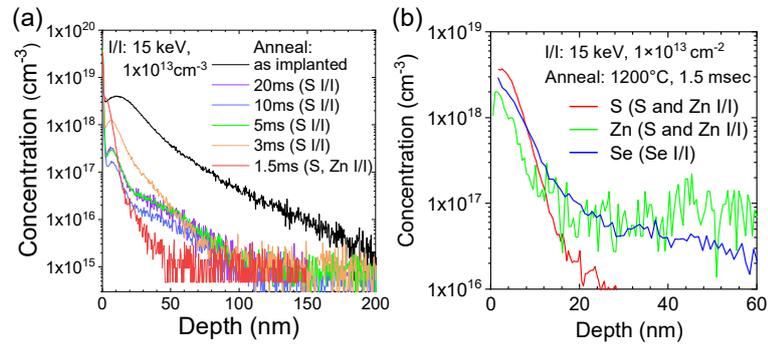

**Fig. 1.** S depth profiles measured by SIMS of S- and S/Zn-implanted Si(100) substrates at different post-implantation annealing (PIA) times. PIA was performed by flash lamp annealing (FLA) at 1200°C for a range of 1.5 msec to 20 msec. (b) S, Zn, and Se depth profiles measured by SIMS of S/Zn- and Se-implanted Si(100) substrates annealed by FLA at 1200°C for 1.5 msec. S, Zn, and Se ion implantations were conducted with the energy and dose of 15 keV and $1 \times 10^{13}$ cm$^{-2}$, respectively.

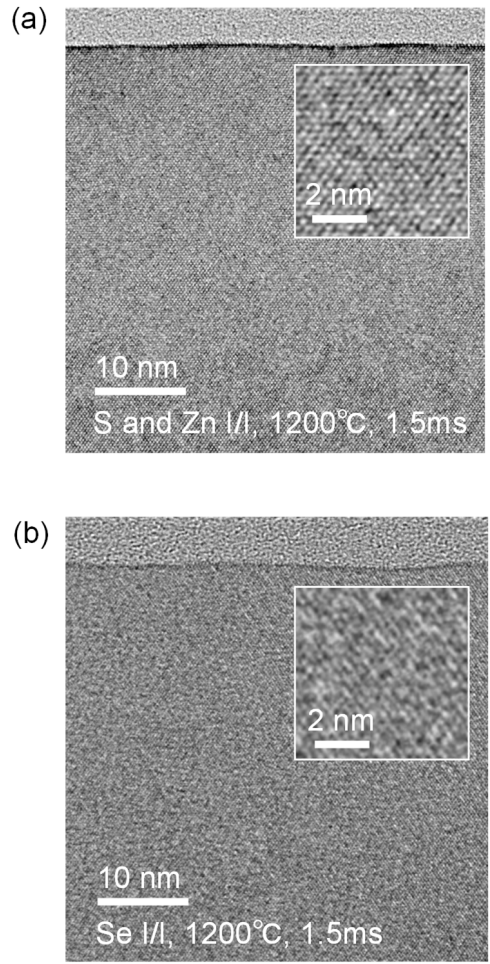

**Fig. 2.** Cross-sectional TEM pictures of (a) S/Zn and (b) Se-implanted Si (100) substrates at a PIA condition of 1200 °C for 1.5 msec. S, Zn, and Se ion implantations (I/I) were conducted with the energy and dose of 15 keV and $1 \times 10^{13} cm^{-2}$.

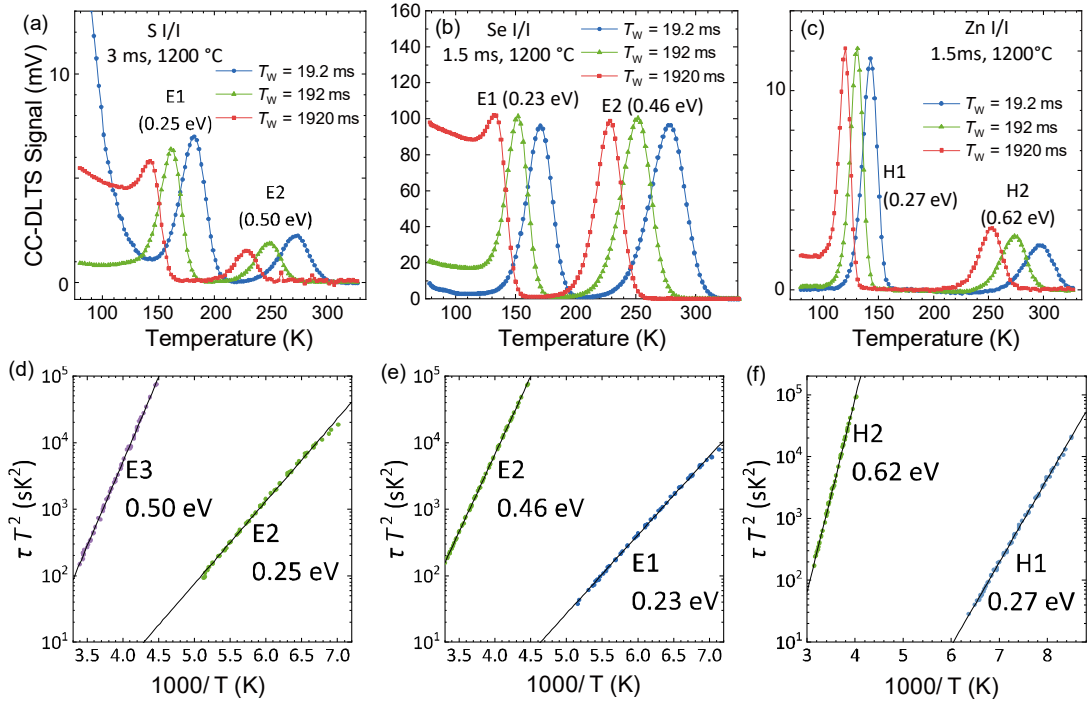

**Fig. 3.** CC-DLTS spectra of (a) S-, (b) Se-implanted n-type, and (c) Zn-implanted p-type Si Schottky diodes with rate windows of 19.2, 192, and 1920 msec. Arrhenius plots obtained from CC-DLTS measurements in (d) S, (e) Se implanted n-type, and (f) Zn implanted p-type Si (100) substrates. S, Se, and Zn ion implantations (I/I) were conducted with the energy and dose of 15 keV and $1 \times 10^{13}$ cm$^{-2}$. PIA condition for S-implanted sample is 1200°C for 3 msec, and Se- and Zn-implanted samples is 1200°C for 1.5 msec by FLA.

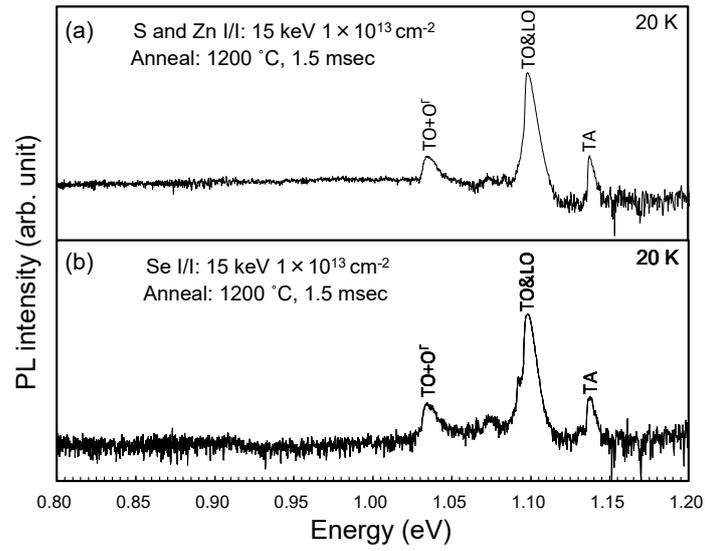

**Fig. 4.** Photoluminescence spectra from (a) S/Zn and (b) Se-implanted Si (100) substrates at a PIA condition of 1200 °C for 1.5 msec. S, Zn, and Se ion implantations (I/I) were conducted with the energy and dose of 15 keV and $1 \times 10^{13}$ cm$^{-2}$.

**Table I.** CC-DLTS analysis results for S-, Se-implanted n-type and (c) Zn-implanted p-type Si (100) samples obtained from the data of Fig. 3.

| Samples | Levels | $E_C - E$ or $E - E_V$ (eV) | $\sigma_n$ or $\sigma_p$ (cm$^2$) | $N_T$ (cm$^{-3}$) |
|---|---|---|---|---|
| S-implanted n-type Si | E1 | 0.25 | 3×10$^{-18}$ | 4.3×10$^{13}$ |
|  | E2 | 0.50 | 4×10$^{-16}$ | 2.3×10$^{13}$ |
| Se-implanted n-type Si | E1 | 0.23 | 5×10$^{-18}$ | 5.1×10$^{14}$ |
|  | E2 | 0.46 | 5×10$^{-17}$ | 7.0×10$^{14}$ |
| Zn-implanted p-type Si | H1 | 0.27 | 8×10$^{-15}$ | 7.7×10$^{13}$ |
|  | H2 | 0.62 | 2×10$^{-14}$ | 2.6×10$^{13}$ |